\documentclass[12pt]{article}
\usepackage{epsf,epsfig,amsmath,amssymb,mathrsfs}
\usepackage{theorem}   
\begin{document}\title{Local Instruction-Set Model for the Experiment of Pan \emph{et al}}

\author{Manuel Aschwanden$^{1,2}$, Walter Philipp$^{2,3,4}$, 
\\ Karl Hess$^{2,5,6}$ and  Salvador Barraza-Lopez$^{2,5}$}

\date{$^{1}$ Integrated Systems Laboratory, Department of Information Technology and Electrical Engineering, ETH, Zurich,
Switzerland \\
$^{2}$ Beckman Institute for Advanced Science and Technology, UIUC\\
$^{3}$ Department of Statistics, UIUC\\
$^{4}$ Department of Mathematics, UIUC\\
$^{5}$ Department of Physics, UIUC\\
$^{6}$ Department of Electrical and Computer Engineering, UIUC\\
 Urbana, Illinois 61801, USA
}


\maketitle
\begin{abstract}
We present a modified local realistic model, based on instruction sets that can be used to approximately reproduce the data of the Pan \emph{et al} experiment. The data of our model are closer to the results of the actual experiment by Pan \emph{et al} than the predictions of their quantum mechanical model. As a consequence the experimental results can not be used to support their claim that quantum nonlocality has been proven.
\end{abstract}

\begin{figure}
    \centering
        \includegraphics[width=1.00\textwidth]{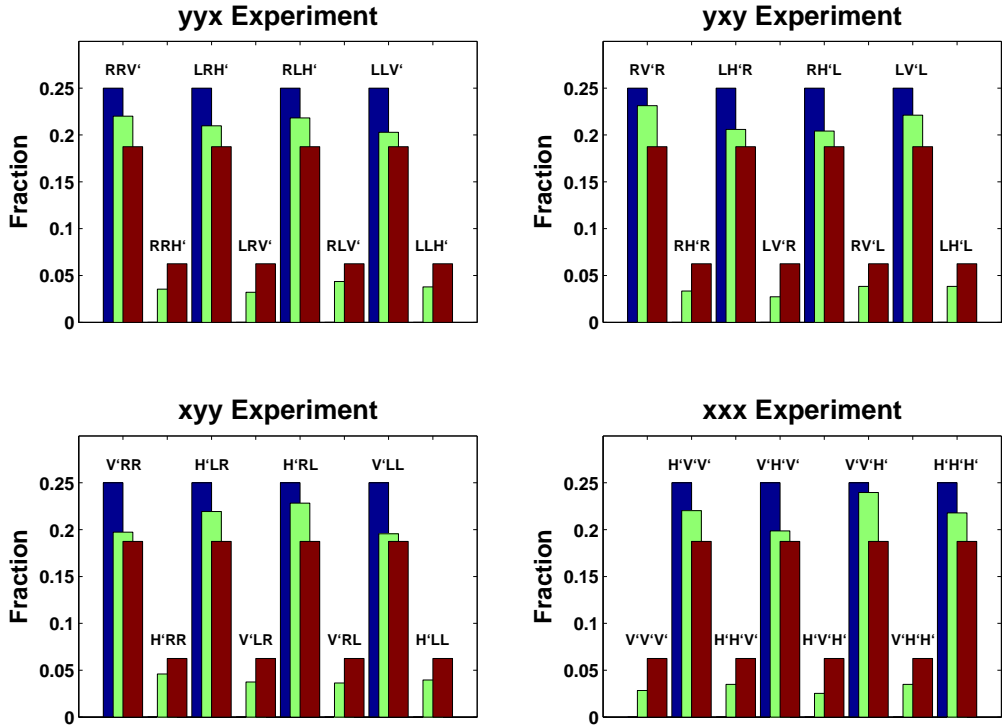}
    \caption{All outcomes for the $yyx$, $yxy$, $xyy$ and $xxx$ experiments. The blue bars represent the quantum mechanical predictions based on the GHZ state, the green bars are the experimental results measured by Pan \emph{et al} and the red bars are the fractions obtained by the instruction sets for the elements of reality shown in Table \ref{tab:ProbabilityDistributionOfElementsOfReality}.}
    \label{fig:fractions}
\end{figure}

The discussions of Einstein-Podolsky-Rosen (EPR) \cite{EPR} and
Bohr \cite{Bohr} relating to the completeness of quantum mechanics
have resulted in significant theoretical \cite{bellbook,CHSC,GHZoriginal} and
experimental \cite{Aspect1,Aspect2, Panetal} work on EPR-type
experiments. At present there is general agreement on the quantum mechanical models (QMM) of the investigated experiments. Nevertheless, several theoretical aspects such as the mathematical nature and properties of parameters used in objective local models\footnote{EPR give the following definition of locality:``Since at the time of measurement the two systems no longer interact, no real change can take place in the second system in consequence of anything that may be done to the first system."\cite{EPR}} \cite{leggett} are still in
discussion. These parameters are considered to represent the physical elements that determine the measurement result in the actual experiments. They may for example characterize quantum particles emitted from a source that is independent from the remote measuring equipment. Or an objective local model may consist of parameters that define the source information and the measuring equipment properties at the time of measurement \cite{hpnp}. In spite of the complexities involved in these discussions, it is universally agreed that as a minimum, an objective local model must at least feature source parameters which are independent of the measuring equipment. These source parameters are considered to be random variables defined on some probability space \cite{feller}. 
\\
The question of the completeness of quantum mechanics is usually addressed by creating EPR-type experiments, for which the quantum mechanical model predicts different, often even opposite results to those obtained from a source and setting dependent objective local model. The measurement results of the actual experiment are then used to compare the two models. Whichever model comes closer to the observed data must be considered to be the better physical model. In the past decades, the Bohm variant of the EPR-Gedankenexperiment \cite{Bohm} and the Greenberger-Horne-Zeilinger multi-particle entanglement Gedankenexperiment \cite{GHZoriginal} were experimentally realized by Alain Aspect \cite{Aspect1, Aspect2} and by Pan \emph{et al} \cite{Panetal} respectively.
\\
In this paper we show that the experimental results reported by Pan \emph{et al} can not be used to decide whether an objective local model \cite{Panetal} or the quantum mechanical model is the better physical model for the observed data. We will use an objective local model based on elements of reality\footnote{EPR define an element of physical reality as follows: ``If, without in any way disturbing a system, we can predict with certainty (i.e., with probability equal to unity) the value of a physical quantity, then there exists an element of physical reality corresponding to this physical quantity."\cite{EPR}} as did Pan \emph{et al}. A minor modification of the assumptions in the objective local model (local realistic model) of Pan \emph{et al} results in a modified local realistic model that gives predictions which are statistically closer to the experimental results  than the predictions of the quantum mechanical model. 
\\
Before introducing our modified local realistic model, we first give a short description  of the work by Pan \emph{et al}\cite{Panetal}. Pan \emph{et al} use the three-photon entangled state also known
as `Greenberger-Horne-Zeilinger' (GHZ) state\cite{GHZoriginal} to experimentally verify quantum nonlocality. 
\\
In the actual experiment, the entangled three-photon GHZ state is analyzed with an experimental setup based on single photons and optical elements. In the Pan \emph{et al} experiment four polarization correlated photons are generated. After having propagated through a setup with several optical elements (wave-plates and beam splitters), the polarizations of three of the four photons, denoted as photon 1, 2 and 3, are measured by equipment consisting of polarization analyzers and three detectors $D_1$, $D_2$ and $D_3$. Whenever detector $D_i$ clicks, it is known that the registered photon $i$ features the polarization indicated by the polarization analyzers in front of the detector. The fourth photon (registered by detector T) is used to guarantee that photon 1, 2 and 3 are in the entangled three-photon GHZ state. Only when all four detectors $T$, $D_1$, $D_2$ and $D_3$ register a photon within a certain time window, it is assumed that the three photons 1, 2 and 3 are in the entangled three-photon GHZ state. This detection of four photons is called fourfold coincidence.
\\
Pan \emph{et al} evaluate the entangled three-photon GHZ state for four different experiments $yyx$, $yxy$, $xyy$ and $xxx$. Here $x$ refers to a photon polarization measurement in the linear polarization basis $H'=+1$/$V'=-1$ ($45^o$/$-45^o$ polarization) and $y$ denotes a measurement in
the circular polarization basis $R=+1$/$L=-1$ (right-handed/left-handed). The $yyx$ experiment means that for photon 1 and 2, the circular polarization and for photon 3 the linear polarization are evaluated. 
\\
The QMM predicts for the entangled three-photon GHZ state four relations that we write symbolically as
\begin{equation}
	yyx=-1\qquad yxy=-1\qquad xyy=-1\label{eqn:1} 
\end{equation}
and 
\begin{equation}
	xxx=+1.\label{eqn:2}
\end{equation}
This means that for each experiment only four of the eight possible polarization
combinations can occur. This quantum mechanical prediction is shown by the blue bars in Figure \ref{fig:fractions}.
\\
In addition to the quantum mechanical analysis, a local realistic model is introduced by Pan \emph{et al}. We quote from their work: ``The only way then for local
realism to explain the perfect correlations predicted by equation
(4) is to assume that each photon carries elements of reality for
both $x$ and $y$ measurements that determine the specific individual
measurement result. For photon $i$ we call these elements of reality
$X_i$ with values $+1$($-1$) for $H'$($V'$) polarizations and $Y_i$ with
values $+1$($-1$) for $R$($L$)..."\cite{Panetal}. Then, the
quantum mechanically predicted relations $Y_1Y_2X_3=-1$, $Y_1X_2Y_3=-1$ and
$X_1Y_2Y_3=-1$ are used by Pan \emph{et al} to restrict the possible
combinations of values for the elements of reality and to find the local realistic prediction for the $xxx$
experiment: ``Because of Einstein locality any specific measurement
for $x$ must be independent of whether an $x$ or $y$ measurement is
performed on the other photon. As $Y_iY_i=+1$, we can write
$X_1X_2X_3=(X_1Y_2Y_3)(Y_1X_2Y_3)(Y_1Y_2X_3)$ and obtain
$X_1X_2X_3=-1$. Thus from a local realist point of view the only
possible results for an $xxx$ experiment are $V'V'V'$, $H'H'V'$,
$H'V'H'$, and $V'H'H'$."\cite{Panetal} Subsequently Pan \emph{et al} show the measurement
results for the $yyx$, $yxy$, $xyy$ and
also the $xxx$ experiments. These measurement results are reproduced by the green bars in Figure \ref{fig:fractions}. 
\\
As it turns out the experimental results \cite{Panetal} do not strictly comply with equation (\ref{eqn:1}). In fact, the measured relative frequency of the states predicted by QM for the GHZ state in equation (\ref{eqn:1}) is only $0.85\pm0.04$, whereas the so-called spurious events do occur with non-negligible relative frequency of $0.15\pm0.02$. The measured relative frequency of the states predicted by QMM for equation (\ref{eqn:2}) is $0.87\pm0.04$ with the spurious events occurring with relative frequency of $0.13\pm0.02$. Because ``the sum of the fractions of all spurious events in the $yyx$, $yxy$, and $xyy$ experiments, that is, $0.45\pm0.03$"\cite{Panetal} is significantly less than $0.87\pm0.04$, the authors interpret their findings as the first three-particle test of local realism following the GHZ argument. In their final analysis they conclude that no objective local model can explain the experimental results and that quantum nonlocality is therefore proven. 
\\However, there exists a logical problem in the assumptions introduced by Pan \emph{et al} for the local realistic model.
Pan \emph{et al} derive their local realistic model based on the predictions of the competing QMM. Before the relations of equation (\ref{eqn:1} are experimentally verified, it is assumed that the local realistic model must also predict these relations. But this procedure conflicts with the basic ideas of a fair scientific comparison between two independent competing models.  
\\
In the following we present a modified local realistic model, based on elements of reality, which produces theoretical predictions that are closer to the results of the actual experiment than the predictions of the QMM. 
\\
We use instruction sets carried by the particles from the common source and therefore still guarantee locality while at the same time extending the model of Pan \emph{et al}. These instruction sets are given in Table \ref{tab:ProbabilityDistributionOfElementsOfReality}. We only list the combinations of elements of reality that are assigned a non-zero probability. For example the instruction set $(H'_1R_1|H'_2R_2|H'_3L_3)$ means that the photon going toward detector $D_1$ (photon 1) has the values $X_1=H'_1$ and $Y_1=R_1$ for the two elements of reality representing the linear and the circular polarization of that photon respectively. Photon 2 has $X_2=H'_2$ and $Y_2=R_2$ and photon 3 has $X_3=H'_3$ and $Y_3=L_3$. 
\\
It can easily be checked 
\begin{table}
	\centering
		\includegraphics[width=0.60\textwidth]{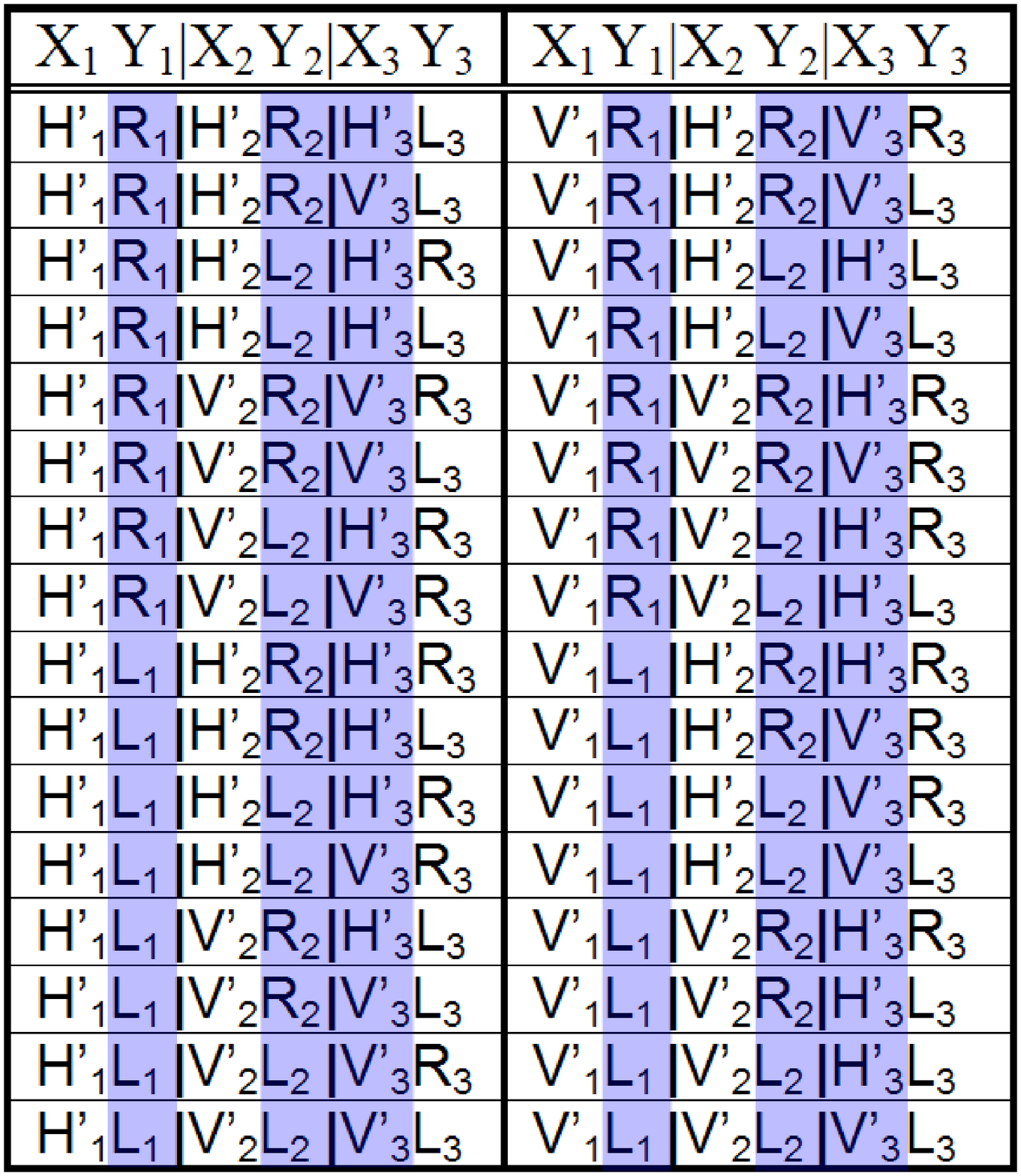}
		\caption{Instruction sets for the elements of reality $X_i$ and $Y_i$. Each of the 32 combinations occurs with equal probability ($\frac{1}{32}$). Shaded entries indicate the relevant elements of reality for the $yyx$ experiment. For example, when the $RRV'$ measurement is
carried out, the following photon triples would result in a
fourfold coincidence count (assuming trigger T clicks):
$(H'_1R_1|H'_2R_2|V'_3L_3)$, $(H'_1R_1|V'_2R_2|V'_3R_3)$,
$(H'_1R_1|V'_2R_2|V'_3L_3)$, $(V'_1R_1|H'_2R_2|V'_3R_3)$,
$(V'_1R_1|H'_2R_2|V'_3L_3)$ and $(V'_1R_1|V'_2R_2|V'_3R_3)$.
Adding the probabilities yields
$\frac{6}{32}$. On the other hand, when the $RRH'$ measurement is
performed, then the photon triples, which would give fourfold
coincidences are $(H'_1R_1|H'_2R_2|H'_3L_3)$ and
($V'_1R_1|V'_2R_2|H'_3R_3$). The sum of the probabilities of
these combinations is $\frac{2}{32}$.}
	\label{tab:ProbabilityDistributionOfElementsOfReality}
\end{table}
that this Table results in a maximum randomness for any individual or two-photon joint measurement. 
\\
If we use the instruction sets of Table
\ref{tab:ProbabilityDistributionOfElementsOfReality} to evaluate
the $yyx$, $yxy$, $xyy$ and $xxx$ experiments, we obtain the
results shown as the red bars in Figure \ref{fig:fractions}.
It can clearly be seen that the predictions of the modified local realistic model are on average closer to the measurement
results of the actual experiment than the quantum mechanical predictions based on the GHZ state. 
\\
We have also computed the results for the $xxy$, $xyx$, $yxx$ and $yyy$ experiments. For all of these experiments it follows from Table \ref{tab:ProbabilityDistributionOfElementsOfReality} that all the eight events (i.e.
$H'_1H'_2R_3$, $H'_1H'_2L_3$,...,$V'_1V'_2L_3$ in the $xxy$
experiment) occur with equal probability $\frac{1}{8}$. This is in agreement with the predictions of the QMM of Pan \emph{et al}.
\\
We are aware that a complete analysis of all aspects of the Pan \emph{et al} experiment may also need to include other observations such as the influence of delays imposed on the photons by the experimental setup. Such delays influence the observed correlations. In general, more elaborate models than the modified local realistic model presented in this paper will be needed to explain all experimental facts. Such models can for example include time and setting dependent equipment parameters \cite{hpnp, Karl}.
\\
In summary we have shown that a modified local realistic model based on elements of reality can explain the actual measurement results reported in \cite{Panetal} with a statistically smaller error than the quantum mechanical model. Therefore, we believe that the Pan \emph{et al} experiment can not be used to draw conclusions about the existence of quantum nonlocality. Moreover, the question whether or not quantum mechanics is a complete theory can not be answered from the reported experimental results because, as we have shown for this particular experiment not even the class of local realistic models based on a small number of instruction sets representing elements of reality can be excluded. 
\\
Future investigations may benefit from a purified version of the Pan \emph{et al} experiment in the sense of \cite{Pan2003}. The use of more sophisticated local hidden variable theories \cite{hpnp, Karl} will also be necessary if the question of quantum nonlocality versus objective local explanations is addressed.
\\
Support of the Office of Naval Research (N00014-98-1-0604) is gratefully acknowledged. MA acknowledges partial funding from IIS (Switzerland) and SB-L acknowledges partial funding from Conacyt (Mexico).

\end{document}